\documentclass[mathleft
]{an}
\usepackage{graphicx}
\usepackage{times}
\begin{document}

\Pagespan{789}{}
\Yearpublication{2006}%
\Yearsubmission{2005}%
\Month{11}%
\Volume{999}%
\Issue{88}%

\title{Elimination of long-term variation from chaotic light curves}

\author{E. Plachy\inst{1}\fnmsep\thanks{\email{eplachy@astro.elte.hu}}, Z. Koll\'ath\inst{2}
}
\titlerunning{Elimination long-term variation from chaotic light curves}
\authorrunning{E. Plachy et al.}
\institute{
Department of Astronomy, E\"otv\"os University, P\'azm\'any P\'eter s\'et\'any 1/A, H-1117, Budapest, Hungary
\and
Konkoly Observatory, MTA CSFK, Konkoly Thege Mikl\'os \'ut 15-17, H-1121, Budapest, Hungary}
\received{}
\accepted{}
\publonline{later}


\abstract{We performed a comparative dynamical investigation of chaotic test data using the global flow reconstruction method. We demonstrate that large-amplitude, long-term variations may have a disturbing effect in the analysis. The Empirical Mode Decomposition method (EMD) and the Fourier filtering were tested to remove the additional variations. Test results show that the elimination of these variations significantly increased the robustness of the reconstructions.
}

\maketitle

\section{Introduction}
Dynamical investigation of chaotic variable stars have already been performed on a few red giant or supergiant variable stars. Most of them belong to the semiregular class (Buchler, Koll\'ath \& Cadmus 2004), but RV Tauri- (Buchler et al. 1996; Koll\'ath et al. 1998) and Mira-type stars (Kiss \& Szatm\'ary 2002) show chaotic behaviour as well. A comprehensive investigation of these classes of stars may recover more candidates. 

Determination of quantitative properties require long and quasi-continuous data set of the light variation. Adequate data is available: the AAVSO and similar databases provide a large collection of visual observations spanning more than hundred years. Complemented with survey observations like ASAS and NSVS data, we can obtain time series that are useful for dynamical studies. 

A suitable analyser tool, the global flow reconstruction method (Buchler \& Koll\'ath 2001) have been used in the studies of the chaotic stars mentioned above (with the exemption of the Mira-type star.) The method is able to detect low dimensional chaos. However, a large-amplitude, long-term variation which is independent of the chaotic dynamics may increase the minimum embedding dimension of the system, reaching the undetectable level. This kind of variation is often visible in observational data, but the origin is not clear in many cases. The so-called long secondary periods, variations over some 10 times the pulsation period, have been observed in several semiregular stars (Woods, Olivier \& Kawaler 2004). Most proposed mechanisms are independent of the pulsation itself, e.g.\ stellar rotation (Kiss et al. 2000), chromospheric activity (Woods et al. 2004), giant cell turnover (Stothers 2010), or binarity (Derekas et al. 2006). RV Tauri stars may also have orbital companions and circumstellar dust disks (Van Winckel et al. 1998). We can not exclude the possibility of observational bias either. The use of different comparison stars may lead to deviations both in the observed average luminosity and the amplitude of the pulsation. The latter is crucial in our analysis, therefore it must be investigated before performing dynamical investigations.

Taking the assumption that long-term variations are not connected to the oscillation of the star, we consider removing these components from the light curves. Nevertheless we believe that detailed tests have to be done before performing such conversion of observational data because chaotic signals may also have low frequency components which, if removed, distort the signal itself. 

Fitting of long-term components is not unambiguous since they are often complex in nature (i.e.\ neither sinusoidal nor smooth trends).
Therefore we tested two different methods to eliminate them: the Empirical Mode Decomposition (EMD) method and the Fourier filtering. The basic difference between the methods is not only the empirical and the analytical approach, but that the EMD process is optimized to handle the time series itself while the latter is to the Fourier transform of it.

We applied the dynamical analysis on a test data with an added long-term variation, and we repeated it on various cleaned data.
In this paper we present the results of this comparison.

\section{Data preparation}

In our previous test paper on global flow reconstruction (Plachy \& Koll\'ath 2010) we presented an analysis of a test data generated by coupling two R\"ossler oscillators. (Equations and parameters can be found in that paper). That data set turned out to be an excellent imitation of chaotic light curves, so we decided to use it in this analysis as well.  

\begin{figure}
\includegraphics[width=83mm]{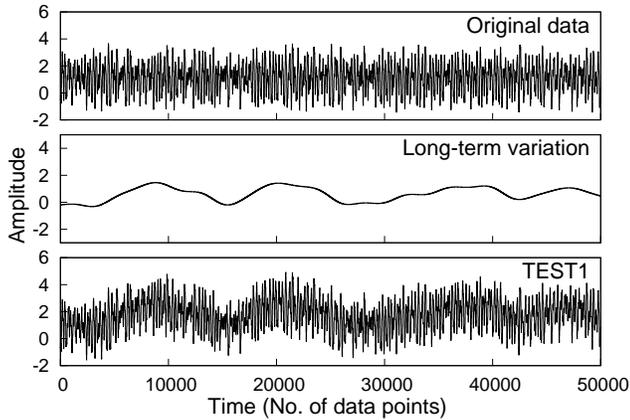}
\caption{Composition of TEST1 data. The original and the long-term variation data are both generated with coupled R\"ossler oscillators (Plachy \& Koll\'ath 2010). TEST1 is the sum of the two. Time is in arbitrary units.}
\label{dat}
\end{figure}

For the long-term variation component we chose a few cycles long fraction of the time series of the chaotic coupled R\"ossler oscillators. It was 
amplified and stretched to the length of the original data. This ensured the long-term variation component to be also chaotic, i.e.\ not constant in frequency and amplitude. TEST1 data set is the simple sum of the R\"ossler and the long-term component (Figure \ref{dat}).

\begin{figure}
\includegraphics[width=80mm]{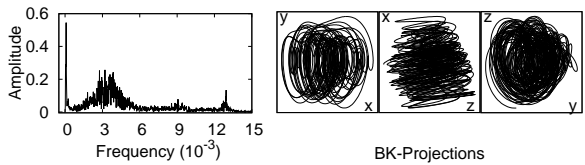}
\caption{Fourier transform and Broomhead-King projections of TEST1. Frequency is in arbitrary units.}
\label{bk}
\end{figure}

TEST2 data was created by removing the extra variation from TEST1 with the EMD method. This method was developed by Huang et al. (1998) and it is used in analysis of nonlinear and nonstationary processes. It is rarely applied in the field of astronomy; however, it showed up in recent studies of solar activity (e.g.\ Vecchio et al. 2012; Li et al. 2012). The decomposition is based on the local characteristic time scale of the data and results in a finite number of intrinsic mode functions (IMFs). The IMFs are amplitude- and frequency-modulated components. Their time scale of variation increases at the successive steps of the decomposition process. This nature of the technique can be used for eliminating long-term variations from the data, by creating an artificial signal as the sum of the first few IMFs. 

\begin{figure}
\includegraphics[width=83mm]{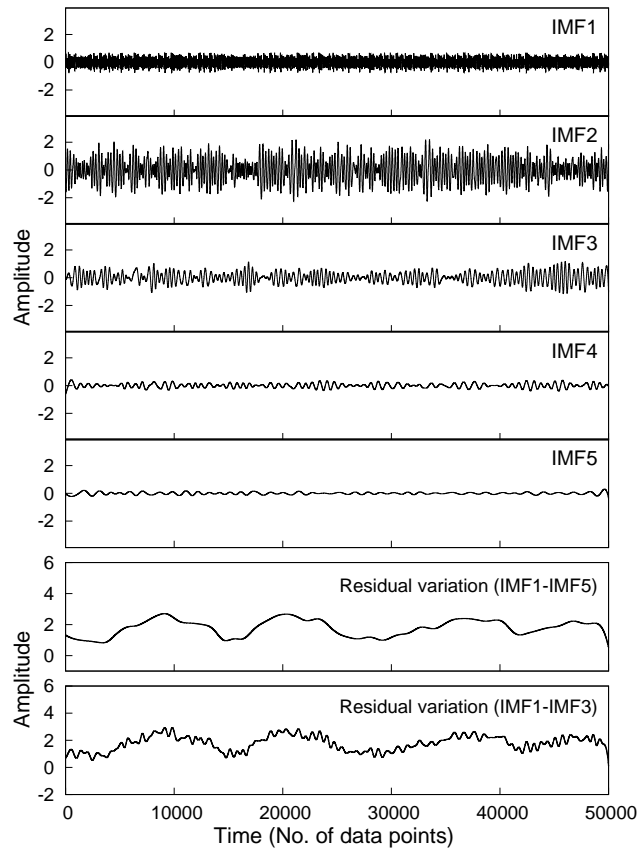}
\caption{The first five intrinsic mode functions (IMFs) of TEST1 data. The lowest two panels show the sums of the residual IMFs, after separating the first five and three IMFs. TEST2 is the sum of the first five IMFs, TEST3 is the sum of the first three IMFs.}
\label{imf}
\end{figure}

The number of required IMFs is not trivial. In our case the sum of the residual IMFs resembled the long-term variation the most, when we used the first five IMFs (Figure \ref{imf}). However, the exact shape of the extra variation is not known in real data, therefore we investigated a different solution as well: TEST3 is the sum of the first three IMFs. 

\begin{figure}
\includegraphics[width=83mm]{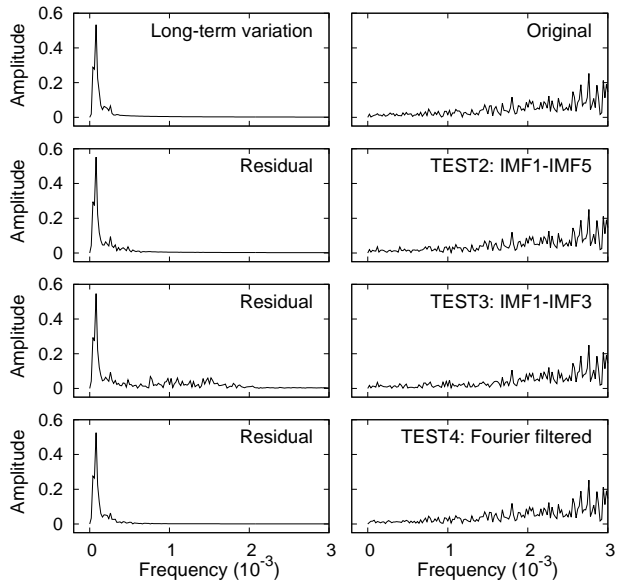}
\caption{Left panels: Comparison of the Fourier-spectra of the extra variation and the variations that were removed in each test. Right panels:
The low frequency region of the original R\"ossler and the different test data.  Frequency is in arbitrary units.}
\label{spectra}
\end{figure}

TEST4 was created using the Fourier filtering technique. We applied a sharp tangent hyperbolic filter function with a midpoint at $5 \cdot 10^{-4}$ frequency value.
We display the low frequency region of Fourier-spectra of TEST2, TEST3, TEST4 and of the removed variations in Figure \ref{spectra}. They are comparable to the original R\"ossler data and the long-term variation we added. If we use less IMFs than it is required, false peaks may appear in the Fourier-spectrum, as in the case of the residual of TEST3. Since the EMD method is optimized in the time domain, the neighbouring IMFs overlap each other in the frequency space and they may contain similar components with opposite phases.

\section{Analysis}
We applied the global flow reconstruction on all four test data. Here we only provide the basic steps of the method and we refer to Buchler \& Koll\'ath (2001) for a detailed description. The input data is converted to so called ``delay vectors", which represent the flow in the phase space. We search for an operator (map) that connects the neighbouring delay vectors. Once the map has been found, we are able to iterate synthetic output data that will be compared to the input. If they are similar, the reconstruction can be considered successful. 

Comparison methods can not be analytical since the data is chaotic. Instead, we use different representations of the data which are compared visually. According the experiences from previous investigations (e.g.\ Buchler et al.\ 2004), the overall shape of the time series, the Fourier Transform, and the Broomhead-King (BK) projection (Broomhead \& King 1987) are useful visualisations, all of which must be similar to the input data. In the case of a successful reconstruction, the quantitative properties of synthetic data are good representations of the original data. 

We determine the Lyapunov exponents and dimension of the synthetic signal. Error estimation is performed with a Monte Carlo-like simulation: we expand the search of maps in the close neighbourhood of the phase space trajectory by adding a small amount of noise to the input data and then applying a cubic spline for smoothing. The delay parameter is also variable in a certain range depending on the sampling of the data. Expansion of the embedding dimension (up to 6D, the largest the dimension that the implementation of the method can handle) gives extra solutions. The following parameter values were set in our tests: noise intensity: 0.001, 0.002, 0.003; smoothing parameter: 0.000, 0.001, 0.002; and time delay: 5-25. The embedding dimensions were 5 and 6.
The outcome of the reconstructions can be diverse: both chaotic and periodic, as well as fix point attractors can be found, but some iterations may also diverge (blow up). The ratio of successful chaotic maps is a good measure of the robustness of the reconstruction.

\section{Results}

\begin{figure}
\includegraphics[width=83mm]{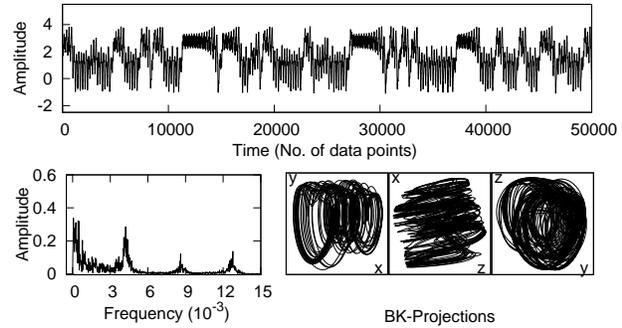}
\caption{An unsuccessful reconstruction of TEST1 data.}
\label{bad}
\end{figure}

\begin{figure}
\includegraphics[width=83mm]{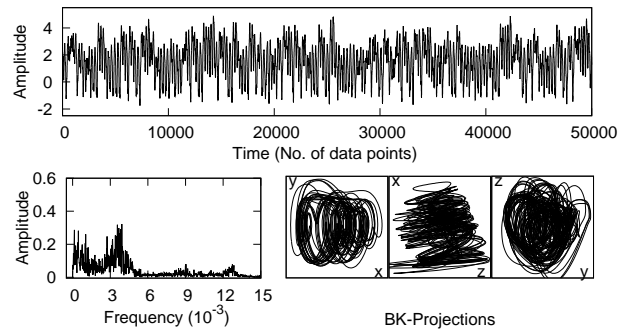}
\caption{An acceptable reconstruction of TEST1 data, showing mid-term variations that are faster than the long-term component in TEST1 (Figure \ref{dat}).}
\label{good}
\end{figure}

The synthetic signals of TEST1 show some sign of extra variation in 40\% of the 5D and in 82\% of the 6D reconstructions. Most of them do not resemble the original, they rather display irregular jumps between two oscillation states. A typical example is shown in Figure \ref{bad}. In the cases where synthetic signals resemble somewhat the original and a long-term variation is also present, it is always shifted to higher frequencies (Figure \ref{good}). Because of these effects, selection of ``good" maps of TEST1 was more difficult than in the other cases.

We summarized the results in Table 1. We listed the number of chaotic maps obtained from the reconstruction, the number of successful chaotic maps with
percentage values in brackets and the average Lyapunov dimension values ($D_L$) of the successful maps. The calculated errors are standard deviations. All Lyapunov dimension values are in agreement with the results of the reconstruction of the original coupled R\"ossler data: $3.03\pm0.30$ in 5D and $3.11\pm0.31$ in 6D values (Plachy \& Koll\'ath 2010). 
After the elimination there was no drastic change in the numbers of chaotic maps, but the number of successful chaotic maps increased significantly.

\begin{table}
\caption{Results of the reconstructions.}
\label{obs}
\begin{tabular}{c c c c c}\hline
 & $D_e$ & Chaotic maps & Successful & $D_L$ \\ 
\hline 
TEST1 & 5D & 61 & 10 (16\%) & $2.92 \pm 0.24$ \\
TEST1 & 6D & 91 & 28 (30\%) & $3.20 \pm  0.17$ \\
\hline
TEST2 & 5D & 90 & 63 (70\%) & $3.06 \pm 0.19$ \\
TEST2 & 6D & 63 & 44 (70\%) & $3.23 \pm  0.10$ \\
\hline
TEST3 & 5D & 99 & 64 (65\%) & $3.05 \pm 0.17$ \\
TEST3 & 6D & 90 & 68 (76\%) & $3.15 \pm  0.18$ \\
\hline
TEST4 & 5D & 82 & 55 (67\%) & $3.01 \pm 0.25$ \\
TEST4 & 6D & 59 & 46 (84\%) & $3.11 \pm 0.19$ \\
\hline
\end{tabular}
\end{table}

\section{Conclusion}
We studied the effects of long-term variation in chaotic data sets with the global flow reconstruction method. The effectiveness of the Empirical Mode Decomposition (EMD) method and the Fourier filtering to remove such variations from the input data was investigated. The number of chaotic synthetic signals we obtained from the reconstructions were similar in all cases, between 140 and 189. However, the percentages of successful chaotic maps show significant differences, making the reconstruction much more robust (three times more successful) if the elimination is applied. The results suggest that removal of high-amplitude, long-term variations can help to detect chaos in light curves, especially if only a few successful maps can be found otherwise. The EMD method was used for the first time for elimination of long-term variation in the context of dynamical analysis with global flow reconstruction. We found that the EMD method is as suitable for such application as the Fourier filtering.

\acknowledgements
Fruitful discussions with L\'aszl\'o Moln\'ar are gratefully acknowledged. The European Union and the European Social Fund have provided financial
support to the project under the grant agreement no. T\'AMOP-4.2.1/B- 09/1/KMR-2010-0003. This work was supported by the Hungarian OTKA grant K83790.

\end{document}